\documentclass[aps, prx, preprint, showpacs, amssymb, amsmath, superscriptaddress]{revtex4-1}
\usepackage{graphicx}
\usepackage{threeparttable}
\usepackage[CJKbookmarks,colorlinks,linkcolor=blue,citecolor=blue]{hyperref}
\usepackage{bm}
\usepackage{times}
\usepackage{array}
\usepackage{float}
\usepackage{colortbl}
\usepackage{multirow}
\usepackage{tabularx}

\begin{document}
\title{First-principles prediction into robust high-performance photovoltaic double perovskites A$_{2}$SiI$_{6}$ (A = K, Rb, Cs)}

\author{Qiaoqiao Li}
\affiliation{State Key Laboratory of Metastable Materials Science and Technology and Key Laboratory for Microstructural Material Physics of Hebei Province, School of Science, Yanshan University, Qinhuangdao 066004, China}

\author{Liujiang Zhou}
\affiliation{Institute of Fundamental and Frontier science, University of Electronic Science and Technology, Chengdu 610054,China}

\author{Yanfeng Ge}
\affiliation{State Key Laboratory of Metastable Materials Science and Technology and Key Laboratory for Microstructural Material Physics of Hebei Province, School of Science, Yanshan University, Qinhuangdao 066004, China}

\author{Yulu Ren}
\affiliation{State Key Laboratory of Metastable Materials Science and Technology and Key Laboratory for Microstructural Material Physics of Hebei Province, School of Science, Yanshan University, Qinhuangdao 066004, China}

\author{Jiangshan Zhao}
\affiliation{Key Laboratory of Soft Chemistry and Functional Materials of MOE, School of Chemical Engineering, Nanjing University of Science and Technology, Nanjing, 210094, China}

\author{Wenhui Wan}
\affiliation{State Key Laboratory of Metastable Materials Science and Technology and Key Laboratory for Microstructural Material Physics of Hebei Province, School of Science, Yanshan University, Qinhuangdao 066004, China}

\author{Kai-Cheng Zhang}
\address{College of Mathematics and Physics, Bohai University, Jinzhou 121013, China}

\author{Yong Liu}
\email{Corresponding author: ycliu@ysu.edu.cn or yongliu@ysu.edu.cn}
\affiliation{State Key Laboratory of Metastable Materials Science and Technology and Key Laboratory for Microstructural Material Physics of Hebei Province, School of Science, Yanshan University, Qinhuangdao 066004, China}


\begin{abstract}
Despite the exceeding 23\% photovoltaic efficiency achieved in organic-inorganic hybrid perovskite solar cells obtaining, the stable materials with desirable band gap are rare and are highly desired. With the aid of first-principles calculations, we predict a new promising family of nontoxic inorganic double perovskites (DPs), namely, silicon (Si)-based halides A$_{2}$SiI$_{6}$ (A = K, Rb, Cs; X = Cl, Br, I). This family containing the earth-abundant Si could be applied for perovskite solar cells (PSCs). Particularly A$_{2}$SiI$_{6}$ exhibits superb physical traits, including suitable band gaps of 0.84-1.15 eV, dispersive lower conduction bands, small carrier effective masses, wide photon absorption in the visible range. Importantly, the good stability at high temperature renders them as promising optical absorbers for solar cells.

\end{abstract}


\maketitle

\section{Introduction}

Since the organic-inorganic hybrid perovskite was first proposed in 2009, the photovoltaic efficiency has been significantly jumped to over 23\% that is close to the maximum efficiency of crystalline silicon solar cell\cite{1, 2, 3}. The hybrid perovskite has a chemical formula ABX$_{3}$, where A$^{+}$ is organic cation (e.g., CH$_{3}$NH$_{3}^{+}$, CH(NH$_{2}$)$_{2}^{+}$), B$^{2+}$ is the post-transition metal with ns$^{2}$ electronic configurations (Pb$^{2+}$, Sn$^{2+}$, Ge$^{2+}$, Sb$^{3+}$, Bi$^{3+}$), and X$^{-}$ is the halide anion (Cl$^{-}$, Br$^{-}$, I$^{-}$). The unique traits, including the ideal direct band gaps, high dielectric constants, shallow defect levels, low electron-hole recombination rates, and long carrier lifetime contribute to the prominent optoelectronic performances of organic-inorganic hybrid perovskites \cite{4, 5, 6, 7, 8, 9, 10}. Despite the remarkable efficiencies of the hybrid lead-based perovskites, as seen in CH$_{3}$NH$_{3}$PbI$_{3}$ (MAPbI$_{3}$) \cite{2, 11, 12, 13, 14, 15, 16}, the poor long-term stability against temperature, oxygen, moisture and exposure to light causes chemical and optical degradation that hinder their wide use \cite{17, 18, 19, 20}. This poor stability is associated with the volatilization and disordered vibrations of small organic cations \cite{18, 19, 21}. Moreover, the toxicity of water-soluble lead compounds drives the exploitation of alternative inorganic lead-free halide perovskite materials with improved stability \cite{22}.

Recently, the inorganic DPs (A$_{2}$B(\uppercase\expandafter{\romannumeral1})B(\uppercase\expandafter{\romannumeral3})X$_{6}$, A$_{2}$B(\uppercase\expandafter{\romannumeral4})X$_{6}$) have been proposed as environmentally friendly and promising alternatives for lead-free hybrid perovskites \cite{23}. However, the light-absorbing materials with suitable and direct band gaps are extremely scarce. For A$_{2}$B(\uppercase\expandafter{\romannumeral1})B(\uppercase\expandafter{\romannumeral3})X$_{6}$ type, Volonakis et al. have performed a theoretical screening on Cs$_{2}$B(\uppercase\expandafter{\romannumeral1})B(\uppercase\expandafter{\romannumeral3})X$_{6}$, B(\uppercase\expandafter{\romannumeral1}) = Cu, Ag, Au, B(\uppercase\expandafter{\romannumeral3}) = Bi, Sb, some of which a portion of materials with appropriate but indirect band gap are predicted \cite{24}. Zhao et al. also screened 64 compounds and only 5 potential direct-band-gap light absorbing materials were obtained\cite{25}. The same type of Cs$_{2}$AgBiBr$_{6}$ has been synthesized experimentally and achieved 2.43\% PCE \cite{26}, the low efficiency may be associated with the large indirect band gap of about 2 eV \cite{27}. For A$_{2}$B(\uppercase\expandafter{\romannumeral4})X$_{6}$ type, there are several materials with suitable band gaps known in theory and experiment, such as Cs$_{2}$TeI$_{6}$, Cs$_{2}$SnI$_{6}$ and Cs$_{2}$TiI$_{6}$ \cite{28, 29, 30}, but only Cs$_{2}$SnI$_{6}$ possesses direct band gap. Therefore, the exploration of inorganic DPs with direct and appropriate band gaps are highly desired. In addition, the fact that crystalline Si solar cells are widely commercialized but more costly in production has inspired us to explore the properties of Si-based DPs compounds in theory.

In this paper, we initially calculated the structure and band gap for a series of earth-abundant Si-based DPs A$_{2}$SiX$_{6}$ (A = K, Rb, Cs, X = Cl, Br, I ) based on the first-principles calculations. Three iodides A$_{2}$SiI$_{6}$ (A = K, Rb, Cs) are found to have suitable direct band gaps of 0.84-1.15 eV. The electronic, optical properties of the three candidates are systemically investigated. The results indicate that the three A$_{2}$SiI$_{6}$ are more suitable for n-type semiconductors and can be utilized as high efficiency optical absorbers. Importantly, K$_{2}$SiI$_{6}$ and Rb$_{2}$SiI$_{6}$ exhibit high thermal stability and thus hold great promise for future optoelectronic devices.

\section{Computational details}

Density hybridized functional theory calculations were performed by the Vienna ab initio simulation package with the projected augmented-wave pseudopotential\cite{31, 32}. The generalized gradient approximation (GGA) with the Perdew-Burke-Ernzerhof (PBE) and that revised for solids (PBEsol) exchange-correlation functional was employed for the structural relaxation\cite{33, 34, 35}. The convergence criteria of the total energy and Hellmann-Feynman force on atom were set to 1$\times$10$^{-5}$ eV and 0.001 eV/{\AA} and the cut-off energy for the plane-wave basis was set to be 400 eV. In order to avoid the underestimates of the band gaps of semiconductors, the Heyd-Scuseria-Ernzerhof (HSE06) functional, which incorporates 25\% Hartree-Fock exchange with a screening parameter of $\omega$= 0.11 bohr$^{-1}$ in addition to 75\% exchange-correlation from the PBE hybrid functional, was adopted to correct the electronic and optical properties \cite{36}. Three-dimensional k-meshes were generated using the Gamma 5$\times$5$\times$5 scheme on electronic and optical calculations \cite{37}. For more accuracy, NBANDS is set to three times the number of valence bands in optical property calculations. For transition dipole moment in optical property calculations, the data were post-processed by VASPKIT code \cite{38}. For chemical bonding interaction analysis, the LOBSTER package was used and the ``pbevaspfit2015'' basis sets including 3\emph{s} and 3\emph{p} orbitals for silicon, and the 5\emph{s} and 5\emph{p} orbitals for iodine and carbon, and the 3\emph{s}, 3\emph{p}, 4\emph{s}, 4\emph{p}, 5\emph{s}, 5\emph{p}, 6\emph{s} orbitals for the metals were taken \cite{39}.

\section{Main results and Discussions}

\begin{figure*}[htbp]
    \centering
    \scalebox{0.60}{\includegraphics{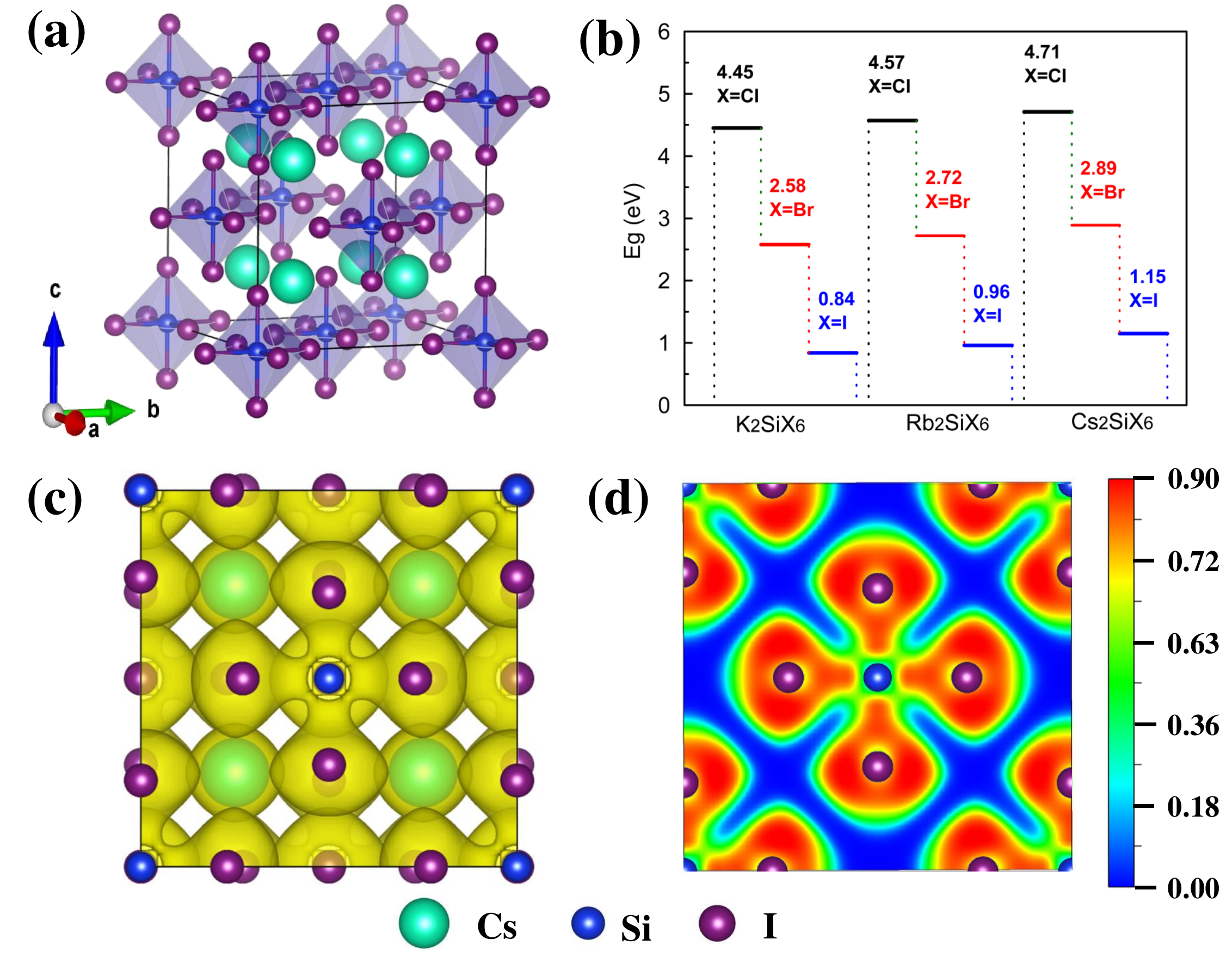}}
       \caption{\label{1}The crystal structure of Cs$_{2}$SiI$_{6}$ with space group Fm$\overline{3}$m. (b) The band gaps of A$_{2}$SiX$_{6}$ (A=K, Rb, Cs; X=Cl ,Br ,I) based on HSE06 functional. (c) The total electronic charge density of Cs$_{2}$SiI$_{6}$ that are viewed from (1 0 0) planar. The isosurface level is 0.5 eV {\AA}$^{-3}$. (d) ELF map sliced of (1 0 0) planar of Cs$_{2}$SiI$_{6}$.}
     \end{figure*}

DP is a defect variant for typical perovskite ABX$_{3}$ structure such as MAPbI$_{3}$\cite{40}. In contrast, for the DP with the general formula A$_{2}$BX$_{6}$, half of the octahedral B-sites atoms are missing, generating nearly isolated octahedral [BX$_{6}^{2-}$] units and presenting as the cubic Fm$\overline{3}$m phases\cite{28}. Represented by Cs$_{2}$SiI$_{6}$, the DP structure is shown in Fig. \ref{1}(a). We can see that the vacancy-ordered DPs formed by face-centered [SiI$_{6}^{2-}$] units and A-site Cs cations uniformly occupy the voids outer the octahedrons. In our calculations, structural parameters of a series of optimized A$_{2}$SiX$_{6}$ DPs (A=K, Rb, Cs; X=Cl, Br, I) are provided in Table S1 (see Ref. \cite{56}). It is found that the lattice constants and Si-X bond lengths gradually increase along with the atomic number of A-site cation and X-site anion severally.

Because there are no related studies on Si-based DPs, and HSE06 hybridized functional predicting band gaps of A$_{2}$B(\uppercase\expandafter{\romannumeral4})X$_{6}$ type compounds are in good agreement with their experimental values, such as Cs$_{2}$SnI$_{6}$, Cs$_{2}$TiI$_{6}$\cite{30, 41}. Therefore, HSE06 functional is adopted for subsequent electronics and optics investigations.

The band gaps of A$_{2}$SiX$_{6}$ (A=K, Rb ,Cs; X=Cl ,Br ,I) DPs are all direct types. Figure \ref{1}(b) displays their band gap values, range from 4.71 to 0.84 eV, including three small gaps with I-containing compounds, namely, K$_{2}$SiI$_{6}$, Rb$_{2}$SiI$_{6}$ and Cs$_{2}$SiI$_{6}$, which are 0.84 eV, 0.96 eV and 1.15 eV, respectively. For compounds that have the same A-site cation, the gaps follow a tendency of A$_{2}$SiCl$_{6}$ $>$ A$_{2}$SiBr$_{6}$ $>$ A$_{2}$SiI$_{6}$ that is consistent with the trend of MAPbI$_{3}$ (X = Cl, Br ,I) \cite{42}. This trend can be analyzed by the electronegativity and density of electronic states (DOS). We present the charge density of  Cs$_{2}$SiI$_{6}$ in Fig. \ref{1}(c). As shown, the charge density mainly distribute over I atoms, while Si atoms has few, indicating charges transfer from the less electronegative Si to the more electronegative I. The overlap of the orbitals along the bonding axis reveals its $\sigma$ bonding type. And taking the DOS of Cs$_{2}$SiX$_{6}$ as an example (see Ref. \cite{56}), valence band edges mainly origin from X atoms, while both X and Si atoms contribute to conduction band edges in three cases. According to Pauling electronegativity\cite{43}, halogen atoms have the strength order of I (2.66) $<$ Br (2.96) $<$ Cl (3.16). The higher electronegativity of X element, the stronger bond interaction between X and Si atoms, thus raising the conduction band and generating larger band gap. Whereas in the same X-site anion situation, the gaps form a tendency of Cs$_{2}$SiX$_{6}$ $>$ Rb$_{2}$SiX$_{6}$ $>$ K$_{2}$SiX$_{6}$ and the reason will be explored later in detail.

To achieve deeper insight into the bonding nature, we analysed the electron localization function (ELF). The ELF renormalizes the values range from 0.00 to 1.00. And the values of 1.00 and 0.50 characterize fully localized and delocalized electrons, respectively, while 0.00 denotes a very low charge density\cite{44}. As displayed in Fig. \ref{1}(d), the large red region around I, corresponding to values about 0.90, implies the dominated localized features of valence electrons. Although limbic region of the Si-I bond is in green, it demonstrates weakly delocalized behavior of their high-energy orbital valence electrons.

According to the Shockley-Queisser limit\cite{45}, which is utilized to evaluate the theoretical photovoltaic conversion efficiency (PCE) in a single junction solar cell, a superior light absorber should possess a band gap ranging from 1.0 to 1.5 eV so as to idealize the efficiency\cite{35}. In tandem devices, the maximum PCE requires a rear cell with a gap of 0.9 to1.2 eV\cite{46}. Therefore, we unify the ideal band gap to be in the range of 0.9-1.5 eV. Considering the slight deviation of the band gap of K$_{2}$SiI$_{6}$ (0.84eV), we here have screened three candidates, K$_{2}$SiI$_{6}$, Rb$_{2}$SiI$_{6}$ and Cs$_{2}$SiI$_{6}$ (A$_{2}$SiI$_{6}$) DPs that can be applied in single or tandem solar cells. Moreover, the spin-orbit coupling (SOC) effect is further considered to correct the band. The corrected values are 0.71 eV, 0.82 eV, and 0.99 eV for K$_{2}$SiI$_{6}$, Rb$_{2}$SiI$_{6}$, and Cs$_{2}$SiI$_{6}$, respectively. These effects on band gaps are much smaller than that of Cs$_{2}$SnI$_{6}$ or Cs$_{2}$TeI$_{6}$ due to the quite lighter B-site element\cite{41}.

\begin{figure*}[htbp]
    \centering
     \scalebox{0.20}{ \includegraphics{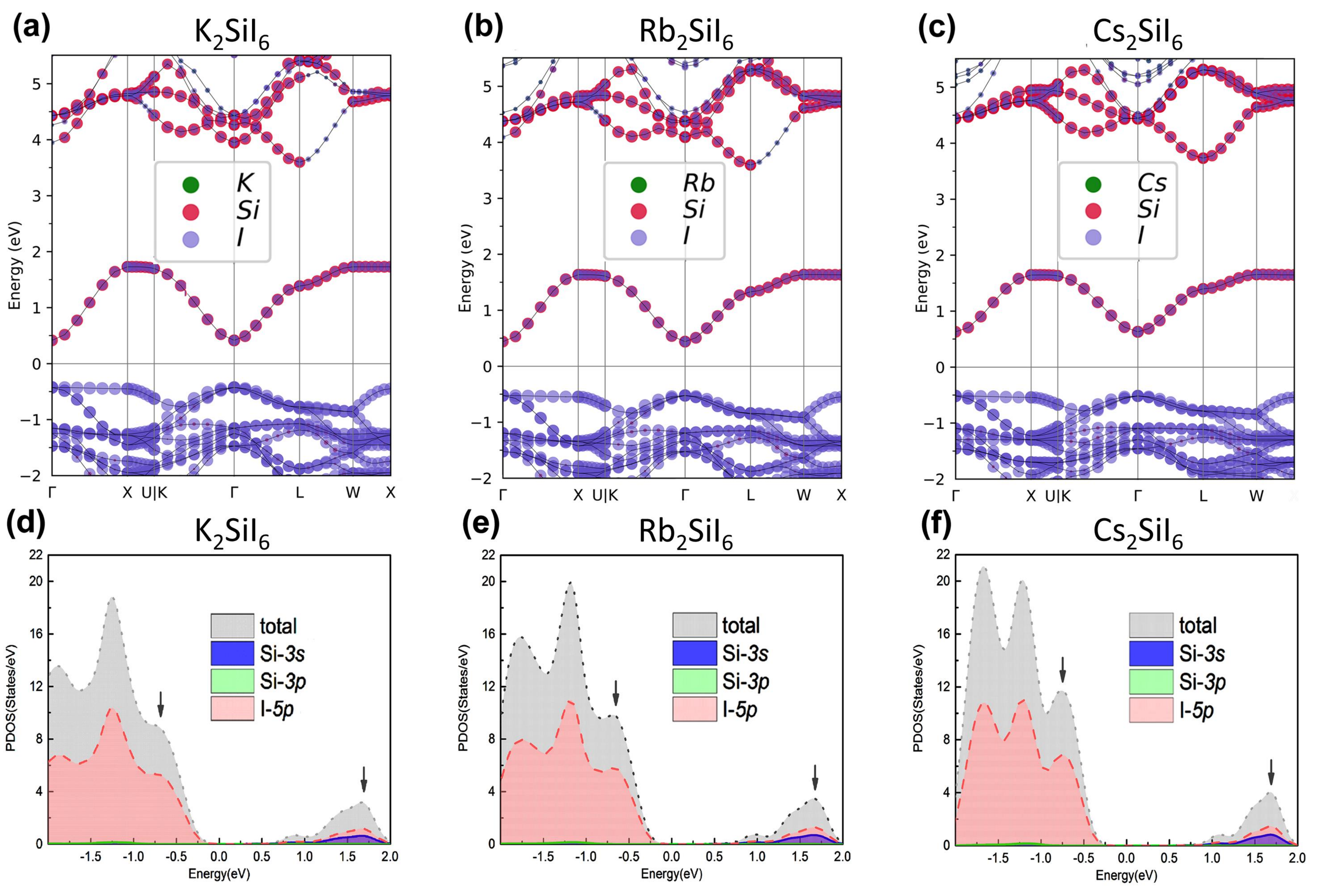}}
       \caption{\label{2}(a)-(c) The projected energy band structures of A$_{2}$SiI$_{6}$ (A=K, Rb, Cs) DPs calculated by HSE06 functional. Fermi-level is set as zero. (d)-(f) PDOS for Si and I atoms of A$_{2}$SiI$_{6}$ DPs. }
     \end{figure*}

Figure \ref{2}(a)-\ref{2}(c) show the projected band structures of the three promising I-based DPs. We can see that in all three cases, the main components of the band edges are analogous to the situation when the anion are Cl$^{-}$ and Br$^{-}$. With the A-site elements barely involve in the formation of band edges, the lower conduction bands are dominated by the Si and I elements, and the upper valence bands mostly come from the I elements. Both conduction band minimum (CBM) and valence band maximum (VBM) locate in the $\Gamma$ points, proving direct band gaps in these A$_{2}$SiI$_{6}$ DPs. To visualize the occupied states, Fig. S2(a)-S2(b)\cite{56} provide the wave function distribution plots of Cs$_{2}$SiI$_{6}$ at VBM and CBM in real space, which can further support our results.

To acquire more band information, we analyze the projected density of electronic states (PDOS) of Si and I atoms near Fermi-level. As shown in Fig. \ref{2}(d)-\ref{2}(f), I-5\emph{p} and hybridized Si-3\emph{s} and I-5\emph{p} orbitals separately constitute the majority of valence and conduction band edges in all three cases. The DOS peaks near the Fermi-level of the three I-based DPs have a significant trend of K$_{2}$SiI$_{6}$ $<$ Rb$_{2}$SiI$_{6}$ $<$ Cs$_{2}$SiI$_{6}$, indicating the increasing tendency of I-5\emph{p} bonding states and anti-bonding states between Si-3\emph{s} and I-5\emph{p} orbitals, resulting in the band gap trend of K$_{2}$SiI$_{6}$ $<$ Rb$_{2}$SiI$_{6}$ $<$ Cs$_{2}$SiI$_{6}$.

\begin{figure*}[htbp]
    \centering
     \scalebox{0.65}{ \includegraphics{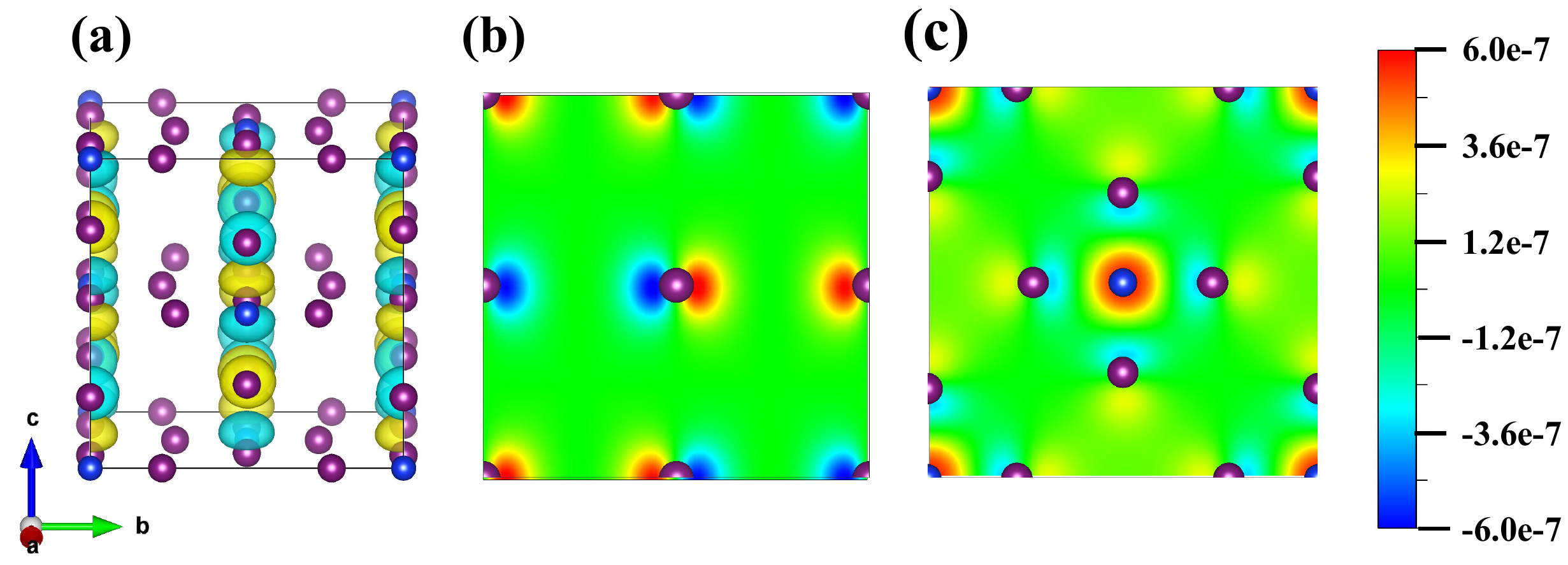}}
       \caption{\label{3}(a) Wave fuction distributions of VBM related to Cs$_{2}$SiI$_{6}$. (b) and (c) are the wave function maps sliced from [1 0 0] direction of VBM and CBM for Cs$_{2}$SiI$_{6}$, respectively. Both slices renormalize the value range from -6.0e$^{-7}$ to 6.0e$^{-7}$, and the common value bar are displayed in the right side. }
     \end{figure*}

It is noticed that above the Fermi level, there exists a well dispersive and isolated band with a bandwidth of 1.32 eV, 1.20 eV and 1.03 eV for K$_{2}$SiI$_{6}$, Rb$_{2}$SiI$_{6}$, and Cs$_{2}$SiI$_{6}$, respectively [Fig. \ref{2}(a)-\ref{2}(c)]. The wide band dispersion indicates the potential favorable electron mobility. Whereas the dispersion of valence bands is weaker than that above conduction bands, which can be typically illustrated by the wave function distributions at VBM and CBM. Taking Cs$_{2}$SiI$_{6}$ as an example, as shown in Fig. \ref{3}(a)-\ref{3}(c), the states of VBM are distributed over partial I atoms, since there are no wave fuction distributions over one third I atoms [Fig. \ref{3}(a)]. But the wave fuction of CBM significantly spreads over all Si and I atoms in our observations. Owing to the smaller absolute value of wave function about CBM [Fig. \ref{3}(b)-\ref{3}(c)], the large delocalized traits related to hybrid Si-3\emph{s} and I-5\emph{p} states are also observed, producing the more dispersive conduction bands than the upper valence bands near the Fermi level.

The well dispersive band directly reflects the small carrier effective mass. By using Equation\ref{eq1}, the carrier effective masses for the K$_{2}$SiI$_{6}$, Rb$_{2}$SiI$_{6}$ and Cs$_{2}$SiI$_{6}$ were calculated around $\Gamma$ points.

\begin{eqnarray}\frac{1}{m^{\ast }}=\frac{1}{h^{2}}\cdot \frac{\partial^{2}E\left ( k \right )}{\partial k^{2}}
\label{eq1}
       \end{eqnarray}

\begin{table}

  \caption{The calculated carrier effective masses for the K$_2$SiI$_6$, Rb$_2$SiI$_6$ and Cs$_2$SiI$_6$ based on HSE06+SOC method. $m_e^*$ $=$ electron effective mass; $m_h^*$ $=$ effective mass of a hole.}
  \label{table1}
  \setlength{\tabcolsep}{9mm}
  \begin{tabular}{lcc}
    \hline
    Compounds & {$m_e^*/m_0$} & {$m_h^*/m_0$}\\
    & $\Gamma-X$ & $\Gamma-Y$\\
    \hline
    K$_2$SiI$_6$ & 0.17 & -0.53 \\
    Rb$_2$SiI$_6$ & 0.18 & -0.56 \\
    Cs$_2$SiI$_6$ & 0.23 & -0.63 \\
    \hline
  \end{tabular}
\end{table}

As their dispersive conduction bands states revealed, all three DPs possess small effective electron masses [Table \ref{table1}], indicating benign conductivity. The heavier hole masses are also in consistence with the prior wave function analysis, reflecting that A$_{2}$SiI$_{6}$ DPs are more suitable for n-type semiconductors. The effective masses of all the three A$_{2}$SiI$_{6}$ DPs are slightly larger than MAPbI$_{3}$\cite{47}, but smaller than other A$_{2}$B(\uppercase\expandafter{\romannumeral4})X$_{6}$ type compounds, such as Rb$_{2}$PtI$_{6}$ and Cs$_{2}$SnI$_{6}$\cite{41}. The electron effective masses of these three DPs have an increasing tendency: K$_{2}$SiI$_{6}$ $<$ Rb$_{2}$SiI$_{6}$ $<$ Cs$_{2}$SiI$_{6}$, which accords with the dispersion degree of their first conduction bands [Fig. \ref{2}(a)-\ref{2}(c)].

The band structure can directly determine the performance of photon absorption. Then the optical properties of the three A$_{2}$SiI$_{6}$ DPs were investigated by calculating the frequency dependent dielectric tensor $\varepsilon \left ( \omega  \right ),\varepsilon \left ( \omega  \right )=\varepsilon _{1}\left ( \omega  \right )+i\varepsilon _{2}\left ( \omega  \right )$, where $\varepsilon _{1}\left ( \omega  \right )$ and $\varepsilon _{2}\left ( \omega  \right )$ are the real and imaginary parts in several, and $\omega$ is the photon frequency\cite{48}. Utilizing dielectric tensor, the optical absorption coefficient $\alpha \left ( \omega  \right )$ can be obtained by the following Equation\ref{eq2}.

\begin{eqnarray}\alpha \left ( \omega  \right )=\frac{\sqrt{2}\omega }{c}\cdot \left [ \sqrt{\varepsilon _{1}\left ( \omega \right )^{2}+\varepsilon _{2}\left ( \omega  \right )^{2}}-\varepsilon _{1}\left ( \omega  \right ) \right ]^{/2}
\label{eq2}
       \end{eqnarray}

\begin{figure}[htbp]
    \centering
    \scalebox{0.095} { \includegraphics{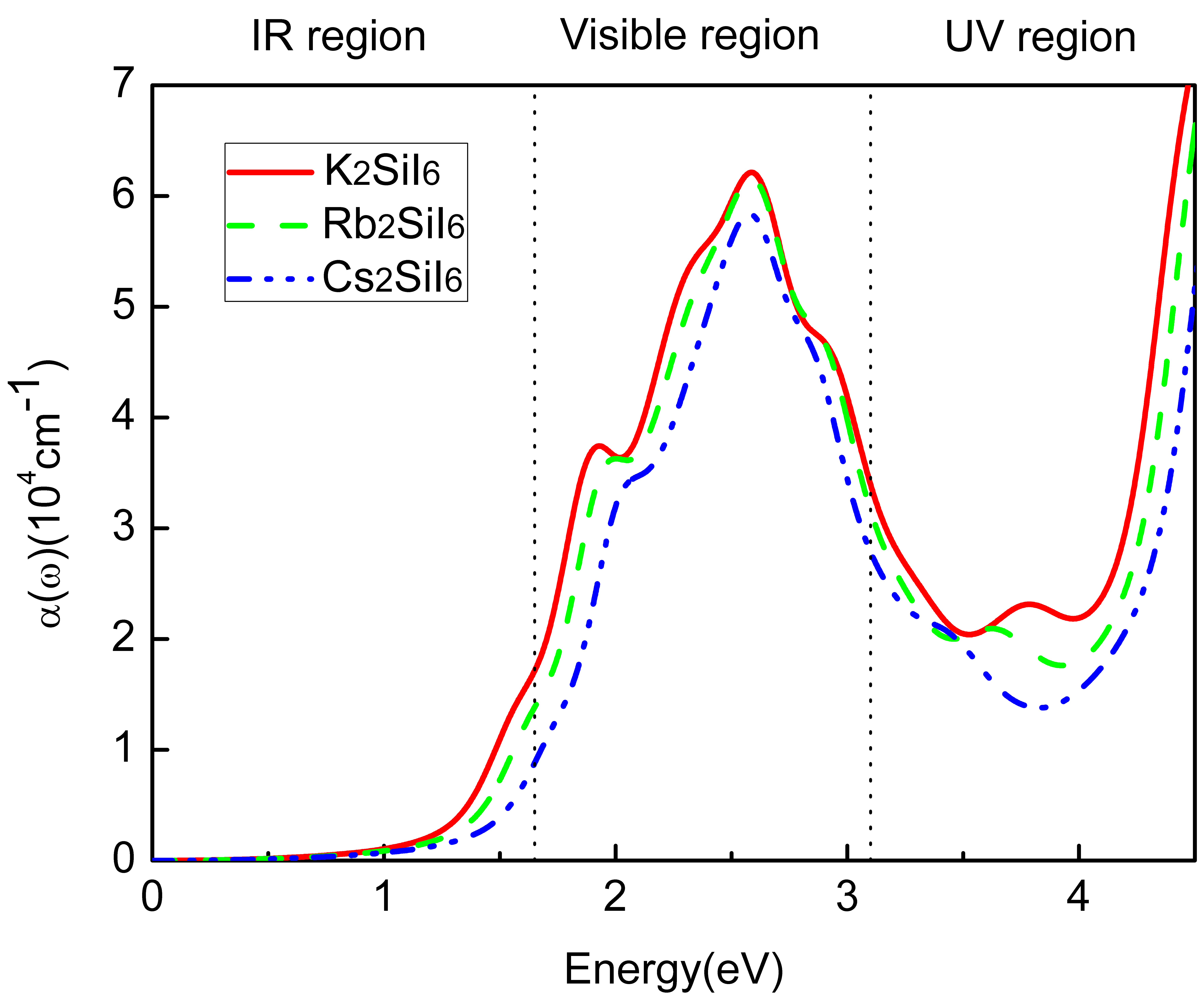}}
       \caption{\label{4}The optical absorption spectra of the three A$_{2}$SiI$_{6}$ DPs calculated by HSE06 functional.}
     \end{figure}

As shown in Fig. \ref{4}, the three A$_{2}$SiI$_{6}$ DPs happen to exhibit an absorption peak in the visible region. With the increase of band gap, the edges of absorption spectra show a blue shift trend. The absorption ability of these three DPs in visible region follows the trend of K$_{2}$SiI$_{6}$ $>$ Rb$_{2}$SiI$_{6}$ $>$ Cs$_{2}$SiI$_{6}$ based on absorption peak and band width. The wide absorptions in visible region are associated to their dispersive lower conduction band and band gap. To evaluate the optical absorption capacity of the three DPs, we have calculated as well the photon absorption coefficients of monocrystalline silicon for the sake of comparison. As shown in Fig. S3\cite{56}, the indirect band gap value of 1.19 eV is consistent with the experimental value of 1.12 eV\cite{3}. The absorption peak of A$_{2}$SiI$_{6}$ in the visible region is lower than that of Si, but the absorption width is significantly wider. Considering crystalline silicon solar cells have achieved efficiency exceeding 25\%\cite{3}, the three A$_{2}$SiI$_{6}$ DPs seem to be promising lead-free perovskite optical absorption layers.

\begin{figure}[htbp]
    \centering
     \scalebox{0.11}{ \includegraphics{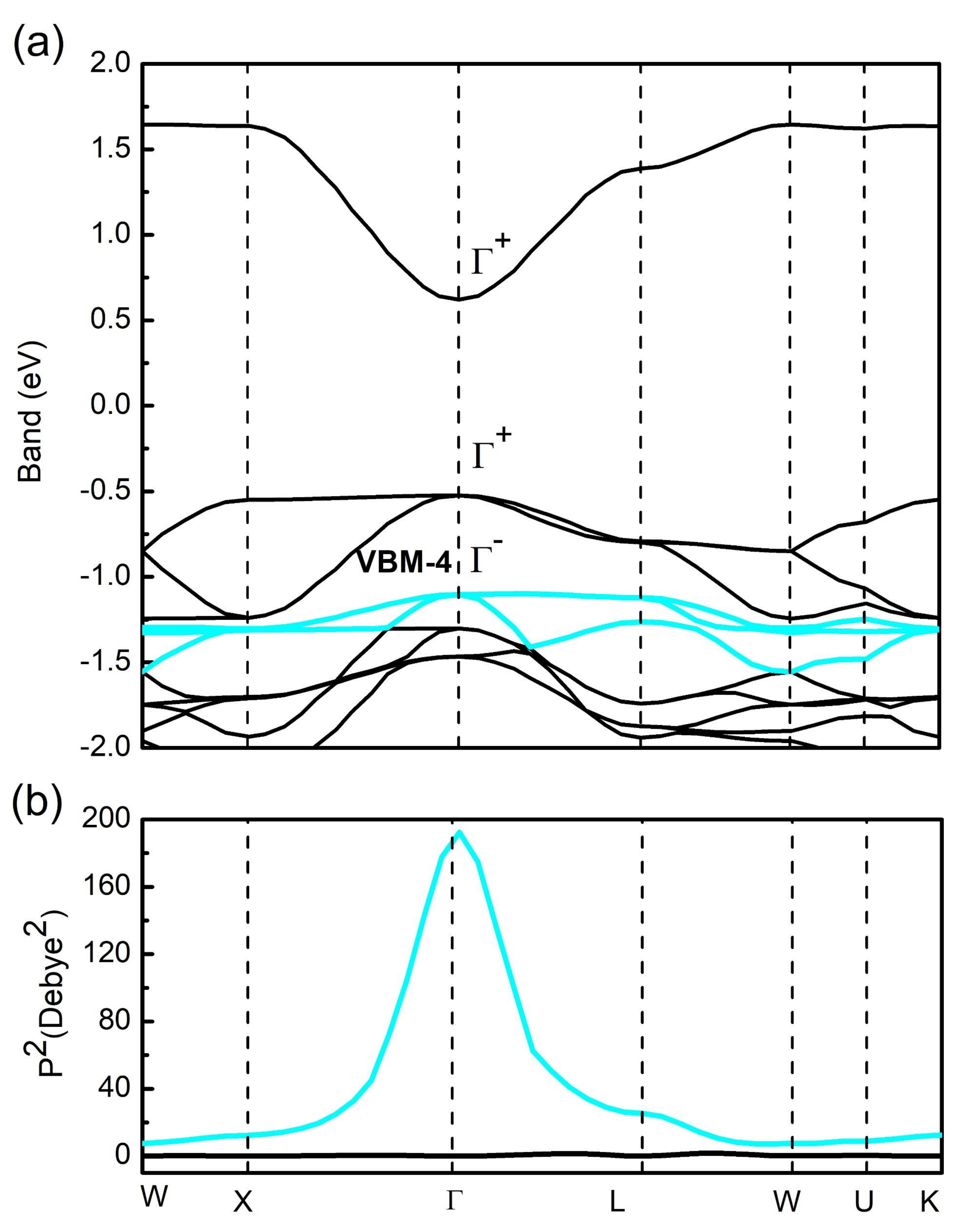}}
       \caption{\label{5}(a) Band structure and parity at $\Gamma$ of Cs$_{2}$SiI$_{6}$, the cyan lines represent the valence bands  corresponding to VBM-4. (b) The cyan line represents the sum of transition matrix elements from the valence band where VBM-4 is located to the conduction band located by CBM, and the black line represents the sum of transition dipole moment from the valence band where VBM is located to the conduction band located by CBM.}
     \end{figure}

To probe the origin of the strong light-harvesting capability of the above three A$_{2}$SiI$_{6}$ DPs, we have analyzed the parity-forbidden transitions via calculating the parity and transition dipole moment between valence bands and the conduction bands of interest. Taking Cs$_{2}$SiI$_{6}$ as an example, the VBM at the $\Gamma$ point exhibits the even parity, possessing a triple degeneracy and the CBM has also even parity [Fig. \ref{5}(a)]. Therefore, the transition from VBM to CBM would not occur and finally produces the zero transition matrix elements. The lower three valence bands below the degenerately top three valence bands are also threefold degenerate and thus are denoted as VBM-4. VBM-4 owns the odd parity, indicating the transition from VBM-4 to CBM is allowed. As a result, there is transition dipole moment from VBM-4 to CBM. Our result reveals that the direct transition from VBM to CBM is forbidden and the transition matrix elements are mainly originated from transition between VBM-4 and CBM. The dipole-moment-allowed direct optical transitions of Cs$_{2}$SiI$_{6}$, K$_{2}$SiI$_{6}$ and Rb$_{2}$SiI$_{6}$ are 1.73 eV, 1.57 eV and 1.63 eV, respectively. In general, according to the distributions of transition matrix elements in Fig. \ref{5}(b), the three A$_{2}$SiI$_{6}$ DPs maintain their absorption characteristics in a wide visible region, suggesting the potential optical absorbers for solar cells. It is worthy to note that K$_{2}$SiI$_{6}$ would achieve the best optical performance among the three DPs due to the best optical transition gap of 1.57 eV according to the Shockley-Queisser standard \cite{45} and the optimum optical absorption in the visible region.

Stability is one of the most important part of judging the application potential of materials. Therefore, we calculated the elastic constants to evaluate mechanical stability of the three I-based DPs. For cubic crystal system, the elastic constants satisfy the Born stability criterion $C_{11}-C_{12}> 0$, $C_{44}> 0$ and $C_{11}+2C_{12}> 0$, manifesting mechanical stability\cite{49}. These elastic constants are defined as

  \begin{eqnarray}C_{ij}=\frac{1}{V_{0}}\cdot \left ( \frac{\partial _{2}E}{\partial\varepsilon _{i}\partial \varepsilon _{j}} \right )
       \end{eqnarray}

Here E is the energy of the crystal, $V_{0}$ denotes equilibrium volume, and $\varepsilon$ gives a strain. Table \ref{table2} lists the elastic constants of the three A$_{2}$SiI$_{6}$ DPs. Our results reveal the mechanical stability of the three A$_{2}$SiI$_{6}$ DPs, since the Born stability criterion are well matched.

\begin{table}
  \caption{Computed elastic constants C$_{11}$, C$_{12}$ and C$_{44}$ of three A$_2$SiI$_6$ DPs}
  \label{table2}
  \begin{tabular}{lcccc}
    \hline
    Compounds & C$_{11}$(GPa) & C$_{12}$(GPa) & C$_{44}$(GPa) & stability \\
    \hline
    K$_2$SiI$_6$ & 13.12 & 3.79 & 5.58 & stable \\
    Rb$_2$SiI$_6$ & 12.94 & 8.09 & 8.59 & stable \\
    Cs$_2$SiI$_6$ & 10.22 & 5.25 & 5.95 & stable \\
    \hline
  \end{tabular}
\end{table}

We also calculated the decomposition enthalpy ($\Delta H$) to evaluate the thermodynamical stability. The $\Delta H$ is defined as

     \begin{eqnarray}\Delta H=E\left ( ASiI_{3} \right )+E\left ( AI_{3} \right )-E\left ( A_{2}SiI_{6} \right )
       \end{eqnarray}

\begin{table*}[htb]
  \caption{The calculated decomposition enthalpy ($\Delta{H}$ in meV/atom) for three A$_2$SiI$_6$ DPs based on PBEsol functional and Bader and Mulliken (in e) Charges in three A$_2$SiI$_6$ DPs using HSE06 functional..}
  \label{table3}
  \setlength{\tabcolsep}{5mm}
  \begin{tabular}{lccccccc}
    \hline
    Compounds & $\Delta{H}$ & \multicolumn{3}{c}{Bader Charge} & \multicolumn{3}{c}{Mulliken Charge}\\
    & & mental & Si & I & mental & Si & I \\
    \hline
    K$_2$SiI$_6$ & 41.11 & 0.89	& 1.26 & -0.508	& 0.89	& 0.20 & -0.330 \\
    Rb$_2$SiI$_6$ & 75.55 & 0.87 & 1.35	& -0.515 & 0.89 & 0.21 & -0.332 \\
    Cs$_2$SiI$_6$ & 68.89 & 0.85 & 1.37	& -0.513 & 0.88	& 0.21 & -0.328 \\
    \hline
  \end{tabular}
\end{table*}

The detailed $\Delta H$ are listed in Table \ref{table3}. The positive values of the three compounds indicate their thermodynamics stability, of which Rb$_{2}$SiI$_{6}$ owns the maximum decomposition energy. The compositional stability can be also demonstrated by the charge transfer between atoms upon its formation. We analyzed the Bader charge and the Mulliken charge\cite{50, 51} and Table \ref{table3} lists the charge transfer of each atom in A$_{2}$SiI$_{6}$. It shows that the metal and Si atoms are positively charged and thus can be regarded as electron donors, while iodine is the only electron acceptor upon the formation of system. Focusing on the amount of charge obtained by iodine in all three cases, Rb$_{2}$SiI$_{6}$ gains the most electrons in the two cases, indicating the optimum combination stability in the three DPs, which is consistent with the result of decomposition energy. In addition, considering the facts that the Pauling electronegativity of Sn (1.96) is stronger than that of Si (1.90)\cite{43} and Cs$_{2}$SnI$_{6}$ has been extensively synthesized in experiments\cite{28, 29, 52, 53}, so it is believed that the three Si-based DPs have the great feasibility of experimental fabrications.

\begin{figure}[htbp]
    \centering
     \scalebox{0.20}{ \includegraphics{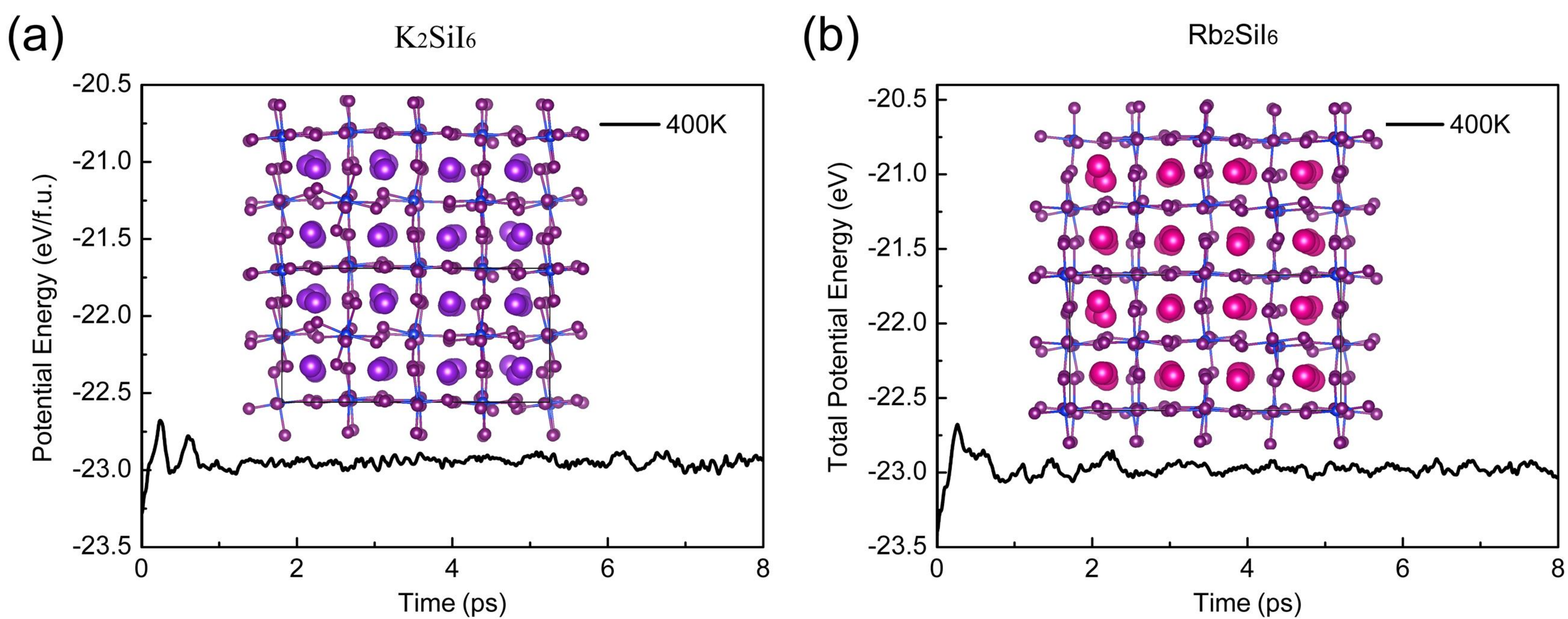}}
       \caption{\label{6}(a) and (b) are the simulated MD potential energy and final structure of K$_{2}$SiI$_{6}$ and Rb$_{2}$SiI$_{6}$ DPs at the temperature of 400 K. }
     \end{figure}

According to international standard (IEC 61646 climatic chamber test), the long-term stability of the 85$^{\circ}$C is required for PSCs. Hence, the molecular dynamics (MD) of the three I-based DPs at the temperature of 400K were simulated. Figure \ref{6} shows the potential energy per formula (f.u.) and the final structures of K$_{2}$SiI$_{6}$ and Rb$_{2}$SiI$_{6}$, which can be seen that the both potential energy finally converges to a range less than 50 meV/atom, indicating outstanding dynamic stability. The well maintained structures of K$_{2}$SiI$_{6}$ and Rb$_{2}$SiI$_{6}$ also prove their stabilities. However, Cs$_{2}$SiI$_{6}$ was found to be dynamically unstable at 400 K [Fig. S4\cite{56}] or at room temperature 300 K, which is similar to Cs$_{2}$AgBiI$_{6}$\cite{54}. So we calculated the integrated crystal orbital Hamilton population (ICOHP) to quantitative description the Si-I bond strength\cite{55}. As demonstrated by Table S2\cite{56}, K$_{2}$SiI$_{6}$ has the maximum value of 1.24, meaning the maximum bonding interactions in Si-X bonds within three DPs. While, Cs$_{2}$SiI$_{6}$ has the minimum value of 1.21. Therefore, the stability problem of Cs$_{2}$SiI$_{6}$ stems from the longer Si-I bonding and the weaker Si-I covalency.

\section{Conclusion}
Using first-principles calculation, we explored and predicted a new kind of Si-based DPs for photovoltaic applications. The results show K$_{2}$SiI$_{6}$, Rb$_{2}$SiI$_{6}$ and Cs$_{2}$SiI$_{6}$ DPs exhibit excellent electronic, optical and stable properties, such as reasonable band gaps, small carrier effective masses, wide photon absorption in visible range, providing more options for the development of lead-free perovskite optical absorbers. These three DPs are more suitable for n-type semiconductors due to their well dispersive lower conduction bands and smaller electron effective masses. K$_{2}$SiI$_{6}$ could achieve the best optical performance in the three DPS due to its best optical transition gap of 1.57 eV and optimum optical absorption in the visible region. According to the comprehensive stability results, Rb$_{2}$SiI$_{6}$ has the best stability, followed by K$_{2}$SiI$_{6}$. Although Cs$_{2}$SiI$_{6}$ has dynamic stability problems, the prospects of the new family of DPs for PCSs are promising.

\begin{acknowledgments}
This work was supported by National Natural Science Foundation of China (No.11904312 and 11904313),the Project of Hebei Educational Department, China(No.ZD2018015 and QN2018012) and the Natural Science Foundation of Hebei Province (No. A2019203507). Thanks to the High Performance Computing Center of Yanshan University.

\end{acknowledgments}

\end{document}